# Particular and unique solutions of DGLAP evolution equation in leading order and gluon structure function at small-$x$

R.Rajkhowa[1] and J. K. Sarma[2]

[1]Physics Department, T. H. B. College, Jamugurihat, Sonitpur, Assam, India

[2.]Physics Department, Tezpur University, Napaam, Tezpur-784 028, Assam, India

[1]E-mail: rasna@tezu.ernet.in, [2] jks@tezu.ernet.in

**Abstract**

We present particular and unique solutions of Dokshitzer- Gribov- Lipatov- Altarelli-Parisi (DGLAP) evolution equation for gluon structure function in leading order (LO) and obtain $t$ and $x$-evolutions of gluon structure function at small-$x$. The results are compared with a recent global parameterization.

## 1. Introduction

In recent papers [1-2], particular and unique solutions of the Dokshitzer- Gribov- Lipatov- Altarelli- Parisi (DGLAP) [3-6] evolution equations for $t$ and $x$-evolutions of singlet and non-singlet structure functions in leading order (LO) and next-to-leading order (NLO) at small-$x$ have been reported. The same technique can be applied to the DGLAP evolution equations for gluon structure function in LO to obtain $t$ and $x$-evolutions of gluon structure function. These LO results are compared with a recent global parameterization [7-8]. Here Section 1, Section 2, and Section 3 will give the introduction, the necessary theory and the results and discussion respectively.

## 2. Theory

The DGLAP evolution equation for gluon structure function has the standard form [9] as

$$\frac{\partial G(x,t)}{\partial t} - \frac{A_f}{t}\left\{\left(\frac{11}{12} - \frac{N_f}{18} + \ln(1-x)\right)G(x,t) + I_g\right\} = 0, \qquad (1)$$

where

$$I_g = \int_x^1 dw \left[\frac{wG(x/w,t) - G(x,t)}{1-w} + \left(w(1-w) + \frac{1-w}{w}\right)G(x/w,t) + \frac{2}{9}\left(\frac{1+(1-w)^2}{w}\right)F_2^S(x/w,t)\right], \qquad (2)$$

$$t = \ln(Q^2/\Lambda^2), \quad A_f = \frac{36}{33-N_f}, \quad N_f \text{ being the number of flavour.}$$

Let us introduce the variable $u = 1-w$ and note that [10]

$$\frac{x}{w} = \frac{x}{1-u} = x \sum_{k=0}^{\infty} u^k. \qquad (3)$$

The series (3) is convergent for $|u|<1$. Since $x<w<1$, so $0<u<1-x$ and hence the convergence criterion is satisfied. Now, using Taylor expansion method [11] we can rewrite $G(x/w, t)$ as

$$G(x/w,t) = G\left(x + x\sum_{k=1}^{\infty} u^k, t\right)$$

$$= G(x,t) + x\sum_{k=1}^{\infty} u^k \frac{\partial G(x,t)}{\partial x} + \frac{1}{2}x^2\left(\sum_{k=1}^{\infty} u^k\right)^2 \frac{\partial^2 G(x,t)}{\partial x^2} + \ldots \qquad (4)$$

which covers the whole range of $u$, $0<u<1-x$. Since $x$ is small in our region of discussion, the terms containing $x^2$ and higher powers of $x$ can be neglected as our first approximation as discussed in our earlier works [1-2, 12-14] and $G(x/w, t)$ can be approximated for small-$x$ as

$$G(x/w,t) \cong G(x,t) + x\sum_{k=1}^{\infty} u^k \frac{\partial G(x,t)}{\partial x}. \qquad (5)$$

Similarly, $F_2^S(x/w, t)$ can be approximated for small-$x$ as

$$F_2^S(x/w,t) \cong F_2^S(x,t) + x\sum_{k=1}^{\infty} u^k \frac{\partial F_2^S(x,t)}{\partial x}. \qquad (6)$$

Using equations (5) and (6) in equations (1) and (2) and performing $u$-integrations we get

$$\frac{\partial G(x,t)}{\partial t} - \frac{A_f}{t}\left[A_1(x)F_2^S(x,t) + B_1(x)\frac{\partial F_2^S(x,t)}{\partial x} + C_1(x)G(x,t) + D_1(x)\frac{\partial G(x,t)}{\partial x}\right] = 0, \qquad (7)$$

where

$$A_1(x) = -\left[\frac{2}{9}(1-x) + \frac{1}{9}(1-x)^2 + \frac{4}{9}\ln x\right],$$

$$B_1(x) = x\left[\frac{4}{9x} + \frac{4}{9}(1-x) + \frac{1}{9}(1-x)^2 + \frac{8}{9}\ln x - \frac{4}{9}\right],$$

$$C_1(x) = \left(\frac{11}{12} - \frac{N_f}{18}\right) + \ln(1-x) - \left[2(1-x) - \frac{1}{2}(1-x)^2 + \frac{1}{3}(1-x)^3 + \ln x\right],$$

$$D_1(x) = x\left[\frac{1}{x} + 2(1-x) + \frac{1}{3}(1-x)^3 + 2\ln x - 1\right].$$

For simplicity we assume [1-2]

$G(x, t) = K(x) F_2^S(x, t)$, where $K(x)$ is a function of $x$. Therefore

$$F_2^S(x,t) = K_1(x)G(x,t), \text{ where } K_1(x) = 1/K(x). \tag{8}$$

Now equation (7) becomes

$$\frac{\partial G(x,t)}{\partial t} - \frac{A_f}{t}\left[P(x)G(x,t) + Q(x)\frac{\partial G(x,t)}{\partial x}\right] = 0, \tag{9}$$

where $P(x) = A_1(x)K_1(x) + B_1(x)\dfrac{\partial K_1(x)}{\partial x} + C_1(x)$ and $Q(x) = B_1(x)K_1(x) + D_1(x)$.

The general solutions of equations (9) is [11, 15] $F(U, V) = 0$, where $F$ is an arbitrary function and $U(x, t, G) = C_1$ and $V(x, t, G) = C_2$ form a solution of equations

$$\frac{dx}{A_f Q(x)} = \frac{dt}{-t} = \frac{dG(x,t)}{-A_f P(x)G(x,t)}. \tag{10}$$

Solving equation (10) we obtain

$$U(x,t,G) = t\exp\left[\frac{1}{A_f}\int\frac{1}{Q(x)}dx\right] \text{ and } V(x,t,G) = G(x,t)\exp\left[\int\frac{P(x)}{Q(x)}dx\right].$$

If $U$ and $V$ are two independent solutions of equation (10) and if $\alpha$ and $\beta$ are arbitrary constants, then $V = \alpha U + \beta$ may be taken as a complete solution of equation (10). Now the complete solution [13-14]

$$G(x,t)\exp\left[\int\frac{P(x)}{Q(x)}dx\right] = \alpha t\exp\left[\frac{1}{A_f}\int\frac{1}{Q(x)}dx\right] + \beta \tag{11}$$

is a two-parameter family of surfaces, which does not have an envelope, since the arbitrary constants enter linearly [11]. Differentiating equation (11) with respect to $\beta$ we get $0 = 1$, which is absurd. Hence there is no singular solution. The one parameter family determined by taking $\beta = \alpha^2$ has equation

$$G(x,t)\exp\left[\int\frac{P(x)}{Q(x)}dx\right] = \alpha t\exp\left[\frac{1}{A_f}\int\frac{1}{Q(x)}dx\right] + \alpha^2. \tag{12}$$

Differentiating equation (12) with respect to $\alpha$, we get $\alpha = -\dfrac{1}{2}t\exp\left[\dfrac{1}{A_f}\int\dfrac{1}{Q(x)}dx\right]$. Putting the value of $\alpha$ in equation (12), we obtain the envelope

$$G(x,t) = -\frac{1}{4}t^2\exp\left[\int\left(\frac{2}{A_f Q(x)} - \frac{P(x)}{Q(x)}\right)dx\right], \tag{13}$$

which is merely a particular solution of the general solution. Now, defining

$$G(x,t_0) = -\frac{1}{4}t_0^2 \exp\left[\int\left(\frac{2}{A_f Q(x)} - \frac{P(x)}{Q(x)}\right)dx\right], \text{ at } t = t_0, \text{ where } t_0 = \ln(Q_0^2/\Lambda^2) \text{ at any lower}$$

value

$Q = Q_0$, we get from equation (13)

$$G(x,t) = G(x,t_0)\left(\frac{t}{t_0}\right)^2, \qquad (14)$$

which gives the *t*-evolution of gluon structure function $G(x, t)$. Again defining,

$$G(x_0,t) = -\frac{1}{4}t^2 \exp\left[\int\left(\frac{2}{A_f Q(x)} - \frac{P(x)}{Q(x)}\right)dx\right]_{x=x_0}, \quad \text{we obtain from equation (13)}$$

$$G(x,t) = G(x_0,t)\exp\left[\int_{x_0}^{x}\left(\frac{2}{A_f Q(x)} - \frac{P(x)}{Q(x)}\right)dx\right] \qquad (15)$$

which gives the *x*-evolution of gluon structure function $G(x, t)$.

For the complete solution of equation (9), we take $\beta = \alpha^2$ in equation (11). If we take $\beta = \alpha$ in equation (11) and differentiating with respect to $\alpha$ as before, we get

$$0 = t\exp\left[\frac{1}{A_f}\int\frac{1}{Q(x)}dx\right] + 1 \text{ from which we can not determine the value of } \alpha. \text{ But if we take } \beta$$

$= \alpha^3$ in equation (11) and differentiating with respect to $\alpha$, we get

$$\alpha = \sqrt{-\frac{1}{3}t\exp\left[\frac{1}{A_f}\int\frac{1}{Q(x)}dx\right]}, \quad \text{from which we get,} \quad G(x,t) = G(x,t_0)\left(\frac{t}{t_0}\right)^{\frac{3}{2}} \text{ and}$$

$$G(x,t) = G(x_0,t)\exp\left[\int_{x_0}^{x}\left(\frac{\frac{3}{2}}{A_f Q(x)} - \frac{P(x)}{Q(x)}\right)dx\right] \text{ as before which are } t \text{ and } x\text{-evolutions}$$

respectively of gluon structure function for $\beta = \alpha^3$.

Proceeding exactly in the same way, we can show that if we take $\beta = \alpha^4$ we get

$$G(x,t) = G(x,t_0)\left(\frac{t}{t_0}\right)^{\frac{4}{3}} \text{ and } G(x,t) = G(x_0,t)\exp\left[\int_{x_0}^{x}\left(\frac{\frac{4}{3}}{A_f Q(x)} - \frac{P(x)}{Q(x)}\right)dx\right] \text{ and so on. So, in}$$

general, if we take $\beta = \alpha^y$, we get

$$G(x,t) = G(x,t_0)\left(\frac{t}{t_0}\right)^{\frac{y}{y-1}} \text{ and } G(x,t) = G(x_0,t)\exp\left[\int_{x_0}^{x}\left(\frac{\frac{y}{y-1}}{A_f Q(x)} - \frac{P(x)}{Q(x)}\right)dx\right], \text{ which give } t$$

and $x$-evolutions respectively of gluon structure function for $\beta = \alpha^y$. We observe if $y \to \infty$ (very large), $y/(y-1) \to 1$.

Thus we observe that if we take $\beta = \alpha$ in equation (11) we can not obtain the value of $\alpha$ and also the required solution. But if we take $\beta = \alpha^2, \alpha^3, \alpha^4, \alpha^5$..... and so on, we see that the powers of $(t/t_0)$ in $t$-evolutions of gluon structure functions are 2, 3/2, 4/3, 5/4….and so on respectively as discussed above. Similarly, for $x$-evolutions of gluon structure functions we see that the numerators of the first term inside the integral sign are 2, 3/2, 4/3, 5/4….and so on respectively for the same values of $\alpha$. Thus we see that if in the relation $\beta = \alpha^y$, $y$ varies between 2 to a maximum value, the powers of $(t/t_0)$ and the numerators of the first term in the integral sign vary between 2 to 1. Then it is understood that the solution of equations (9) obtained by this methodology is not unique and so the $t$ and $x$-evolution of gluon structure function obtained by this methodology is not unique. Thus by this methodology, instead of having a single solution we arrive a band of solutions, of course the range for these solutions is reasonably narrow.

Again, for $Q^2$ values much larger than $\Lambda^2$, the effective coupling is small and a perturbative description in terms of quarks and gluons interacting weakly makes sense. For $Q^2$ of order $\Lambda^2$, the effective coupling is infinite and we cannot make such a picture, since quark and gluons will arrange themselves into strongly bound clusters, namely, hadrons [16]. Also the perturbation series breaks down and structure functions must vanish [17]. Thus, $\Lambda$ can be considered as the boundary between a world of quasi-free quarks and gluons, and the world of pions, protons, and so on. The value of $\Lambda$ is not predicted by the theory; it is a free parameter to be determined from experiment. It should expect that it is of the order of a typical hadronic mass [16]. The value of $\Lambda$ is so small that we can take at $Q = \Lambda$, $F_2^{S'}(x, t) = 0$ due to conservation of the electromagnetic current [18]. Since the relation between gluon and singlet structure function is $G(x, t) = K_1(x)F_2^S(x, t)$, therefore $G(x, t) = 0$ at $Q = \Lambda$. Using this boundary condition in equations (11) we get $\beta = 0$ and

$$G(x,t) = \alpha t \exp\left[\int\left(\frac{1}{A_f Q(x)} - \frac{P(x)}{Q(x)}\right)dx\right]. \tag{16}$$

Now, defining $G(x,t_0) = \alpha t_0 \exp\left[\int\left(\frac{1}{A_f Q(x)} - \frac{P(x)}{Q(x)}\right)dx\right]$, at $t = t_0$, where $t_0 = \ln(Q_0^2/\Lambda^2)$ at any lower value $Q = Q_0$, we get from equation (16)

$$G(x,t) = G(x_0,t)\left(\frac{t}{t_0}\right), \qquad (17)$$

which gives the *t*-evolution of gluon structure function $G(x, t)$ in LO. Again defining,

$G(x_0,t) = \alpha t \exp\left[\int\left(\frac{1}{A_f Q(x)} - \frac{P(x)}{Q(x)}\right)dx\right]_{x=x_0}$, we obtain from equation (16)

$$G(x,t) = G(x_0,t)\exp\left[\int_{x_0}^{x}\left(\frac{1}{A_f Q(x)} - \frac{P(x)}{Q(x)}\right)dx\right], \qquad (18)$$

which gives the *x*-evolution of gluon structure function $G(x, t)$ in LO. We observed that unique solutions (equations (17) and (18)) of DGLAP evolution equation for gluon structure function are same with particular solutions for y maximum in $\beta = \alpha^y$ relation in LO.

## 3. Results and discussion

In the present paper, we present our result of *t*-evolution of gluon structure function qualitatively and compare result of *x*-evolution with a recent global parameterization [7-8]. These parameterizations include data from H1, ZEUS, DO, CDF experiment. Though we compare our results with $y = 2$ and $y =$ maximum in $\beta = \alpha^y$ relation with the parameterization, our result with $y =$ maximum is equivalent to that of unique solution.

In figure 1(a-b), we present our results of *t*-evolutions of gluon structure functions $G(x, t)$ qualitatively for the representative values of *x* given in the figures for $y = 2$ (upper solid and dashed lines) and *y* maximum (lower solid and dashed lines) in $\beta = \alpha^y$ relation. We have taken arbitrary inputs from recent global parameterizations MRST2001 (solid lines) and MRST2001J (dashed lines) in figure 1(a) at $Q_0^2 = 1$ GeV$^2$ [7] and MRS data in figure 1(b) at $Q_0^2 = 4$ GeV$^2$ [8]. It is clear from figures that *t*-evolutions of gluon structure functions depend upon input $G(x, t_0)$ values.

For a quantitative analysis of *x*-distributions of gluon structure functions $G(x, t)$, we calculate the integrals that occurred in equation (15) for $N_f = 4$. In figure 2(a-b), we present our results of *x*-distribution of gluon structure functions for $K_1(x) = ax^b$, where 'a' and 'b' are constants, for representative values of $Q^2$ given in each figure, and compare them with recent global parameterizations [7] for $y$ minimum in the relation $\beta=\alpha^y$. In figure 2(a), we observe that agreement of the results with parameterization is found to be very poor for any values of '$a$' and '$b$' at low-$x$ and agreement is found to be good at high-$x$ at $a = 372$ and $b = 4$ (thick solid line). In figure 2(b), agreement of the results with parameterizations is found to be good at $a = 135$ and $b = 1.8$ (thick solid line) in $\beta = \alpha^y$ relation. In the same figures we present the sensitivity of our results for different values of '$a$' at fixed value '$b$'. Here we take $b = 4$ in figure 2(a) and $b = 1.8$ in figure 2(b). We observe that if value of '$a$' is increased or decreased, the curve goes upward or downward direction respectively. But the nature of the curves is similar. Here thin solid and dotted lines are MRST 2001 and MRST2001J [7] parameterizations.

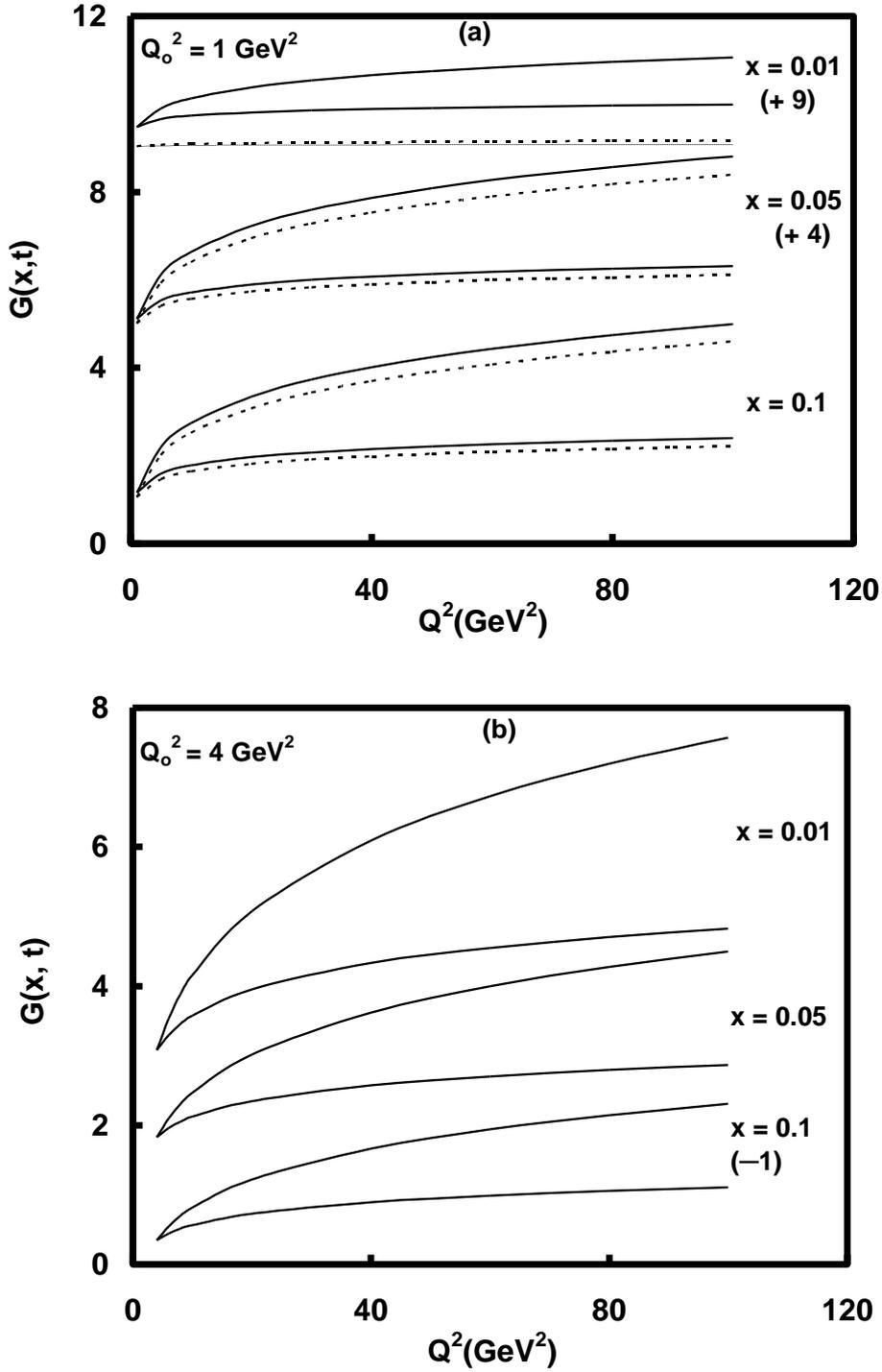

**Figure 1(a-b):** Results of *t*-evolutions of gluon structure functions for the representative values of *x* given in the figures for $y = 2$ (upper solid and dashed lines) and *y* maximum (lower solid and dashed lines) in $\beta = \alpha^y$ relation. We have taken arbitrary inputs from recent global parameterizations MRST2001 (solid lines) and MRST2001J (dashed lines) in figure 1(a) and MRS data in figure 1(b) at $Q_0^2 = 1$ GeV$^2$ and $Q_0^2 = 4$ GeV$^2$ respectively. For convenience, value of each data point is increased by adding 9 and 4 for $x = 0.01$ and $x = 0.05$ respectively in figure 1(a) and decreased by subtracting 1 for $x = 0.1$ in figure 1(b).

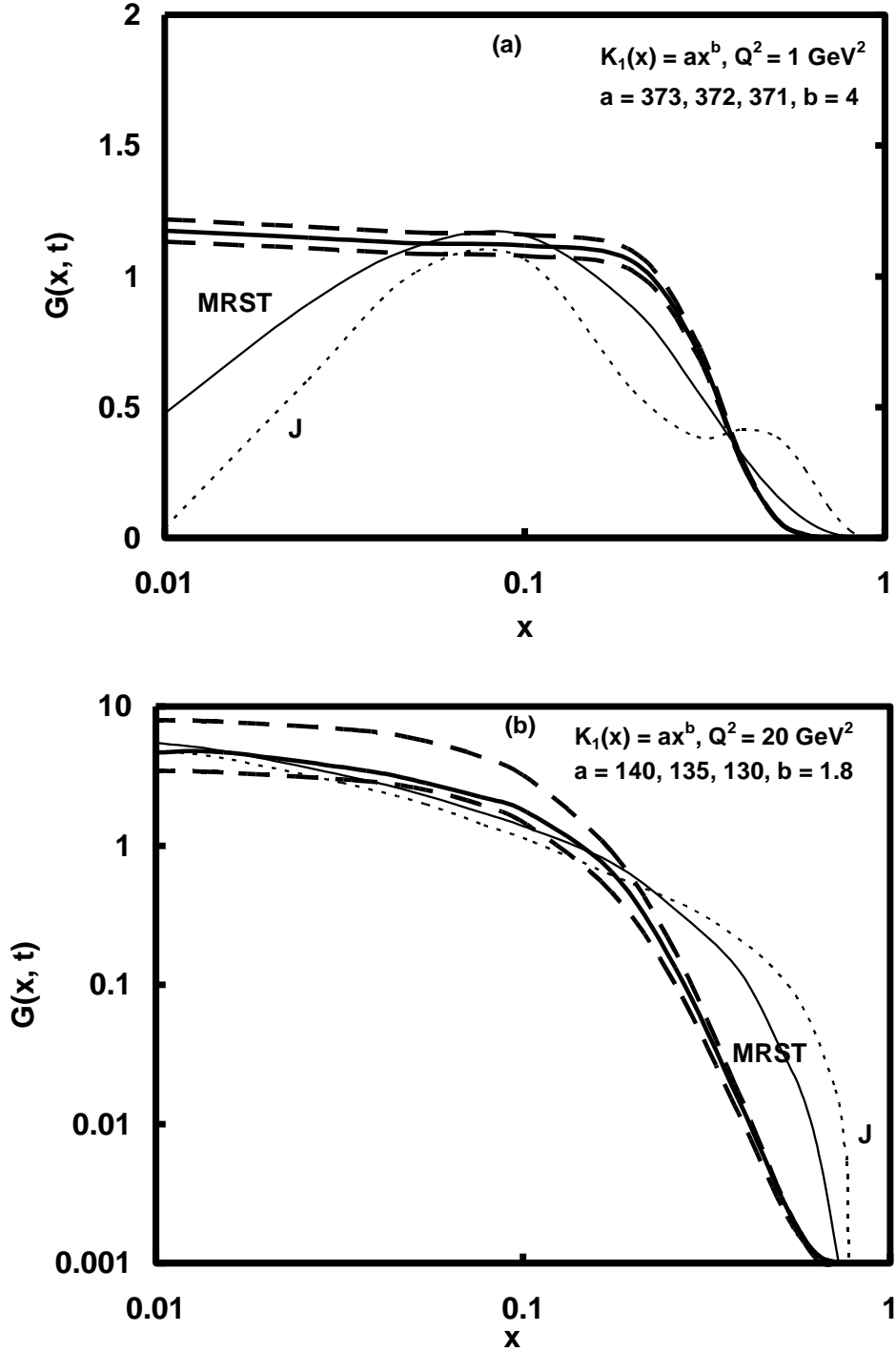

**Figure 2(a-b):** Results of *x*-distribution of gluon structure functions for $K_1(x) = ax^b$, where '*a*' and '*b*' are constants for representative values of $Q^2$ given in each figure, and compare them with recent global parameterizations for *y* minimum (thick solid lines) in the relation $\beta = \alpha^y$. In the same figures we present the sensitivity of our results for different values of '*a*' at fixed value '*b*'. Here we take $b = 4$ in figure 2(a) and $b = 1.8$ in figure 2(b).

In figure 3(a-b), we present the sensitivity of our results for different values of '$b$' at fixed value of '$a$'. Here we take $a = 372$ in figure 3(a) and $a = 135$ in figure 3(b). We observe that, agreement of the results (thick solid lines) with parameterizations is good in figure 3(a) at b = 4 and figure 3(b) at $b = 1.8$. If value of '$b$' is increased or decreased the curve goes downward or upward directions. But the nature of the curve is similar.

In figure 4(a-b), we present our results of $x$-evolution of gluon structure function $G(x, t)$ for $K_1(x) = ax^b$ for $y$ minimum (lower thick solid lines) and maximum (upper thick solid lines) in relation $\beta = \alpha^y$ at same parameter values $a = 372$, $b = 4$ in figure 4(a) and $a = 135$, $b = 1.8$ in figure 4(b) and for representative values of $Q^2$ given in each figure, and compare them with recent global parameterizations [7]. We observe that result of $x$-evolution of gluon structure function for $y$ maximum (long dashed lines) coincide with result of $x$-evolution of gluon structure function for $y$ minimum (lower thick solid lines) when $a = 375$, $b = 4.7$ in figure 4(a) and $a = 134$, $b = 2$ in figure 4(b). That means if $y$ varies from minimum to maximum, then value of parameter '$a$' varies from 372 to 375 and '$b$' varies from 4 to 4.7 in figure 4(a) and '$a$' varies from 135 to 134 and '$b$' varies from 1.8 to 2 in figure 4(b).

In figure 5(a-b), we present our results of $x$-distribution of gluon structure functions $G(x, t)$ for $K_1(x) = ce^{-dx}$, where '$c$' and '$d$' are constants for representative values of $Q^2$ given in each figure, and compare them with recent global parameterizations [7] for $y$ minimum in the relation $\beta = \alpha^y$. In figure 5(a), we observe that agreement of the results with parameterization is found to be very poor for any values of '$c$' and '$d$' at low-$x$ and agreement is found to be good at high-$x$ at $c = 300$ and $d = -3.8$ (thick solid line). In figure 5(b) agreement of the results with parameterizations is found to be good at $c = 5$ and $d = -28$ (thick solid line). In the same figures, we present the sensitivity of our results for different values of '$c$' by thick dashed lines at fixed value '$d$'. Here we take $d = -3.8$ in figure 5(a) and $d = -28$ in figure 5(b).

We observe that if value of '*c*' is increased or decreased, the curve goes upward or downward directions respectively. But the nature of the curve is similar.

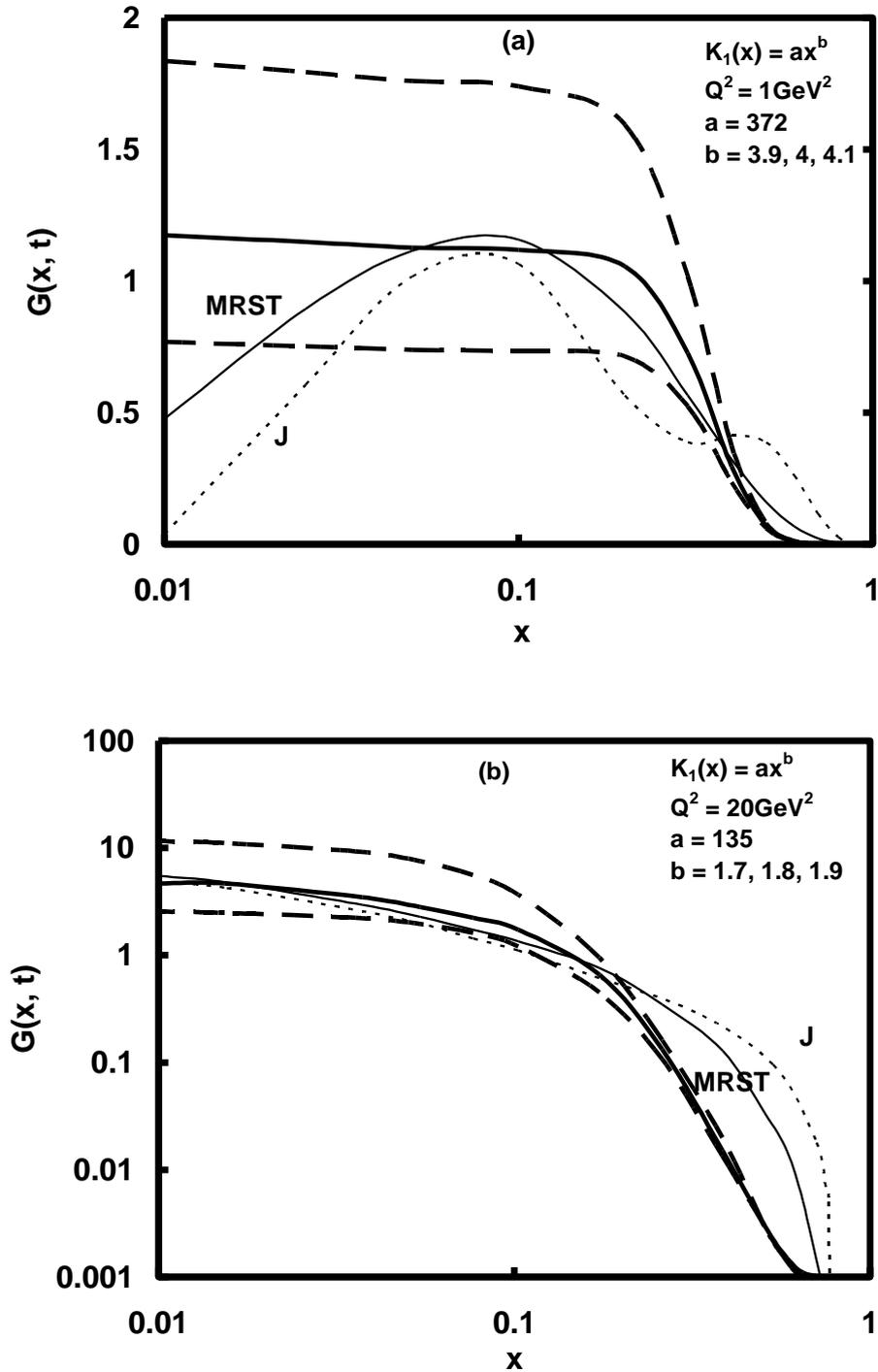

**Figure 3(a-b):** Sensitivity of our results for different values of '*b*' at fixed value of '*a*'. Here we take $a = 372$ in figure 3(a) and $a = 135$ in figure 3(b).

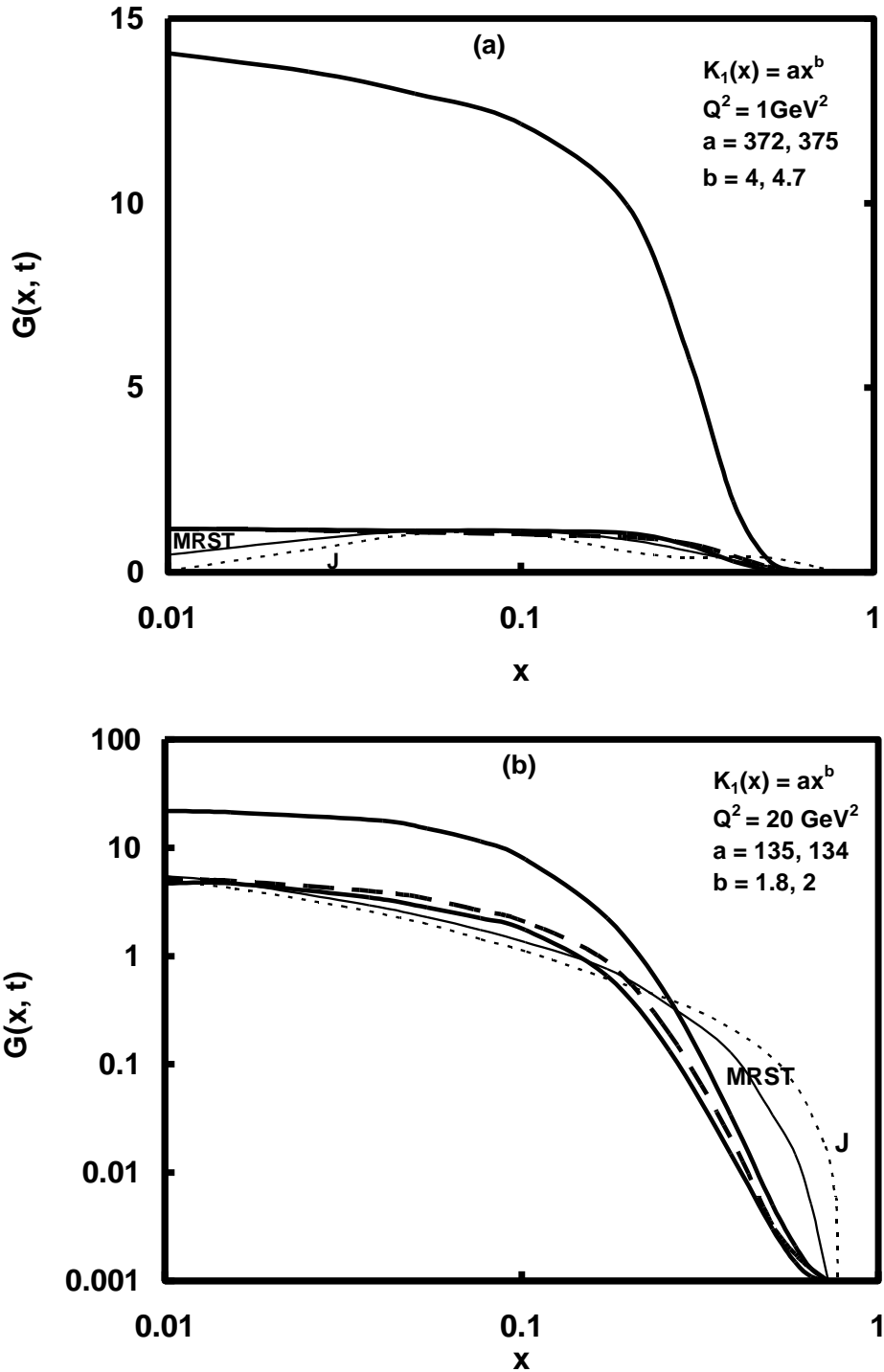

**Figure 4(a-b):** Results of $x$-evolution of gluon structure function $G(x, t)$ for $K_1(x) = ax^b$ for $y$ minimum (lower thick solid lines) and maximum (upper thick solid lines) in relation $\beta = \alpha^y$ at same parameter values $a = 372$, $b = 4$ in figure 4(a) and $a = 135$, $b = 1.8$ in figure 4(b) and for representative values of $Q^2$ given in each figure, and compare them with recent global parameterizations. Result of $x$-evolution of gluon structure function for $y$ maximum (long dashed lines) coincide with result of $x$-evolution of gluon structure function for $y$ minimum (lower thick solid lines) when $a = 375$, $b = 4.7$ in figure 4(a) and $a = 134$, $b = 2$ in figure 4(b).

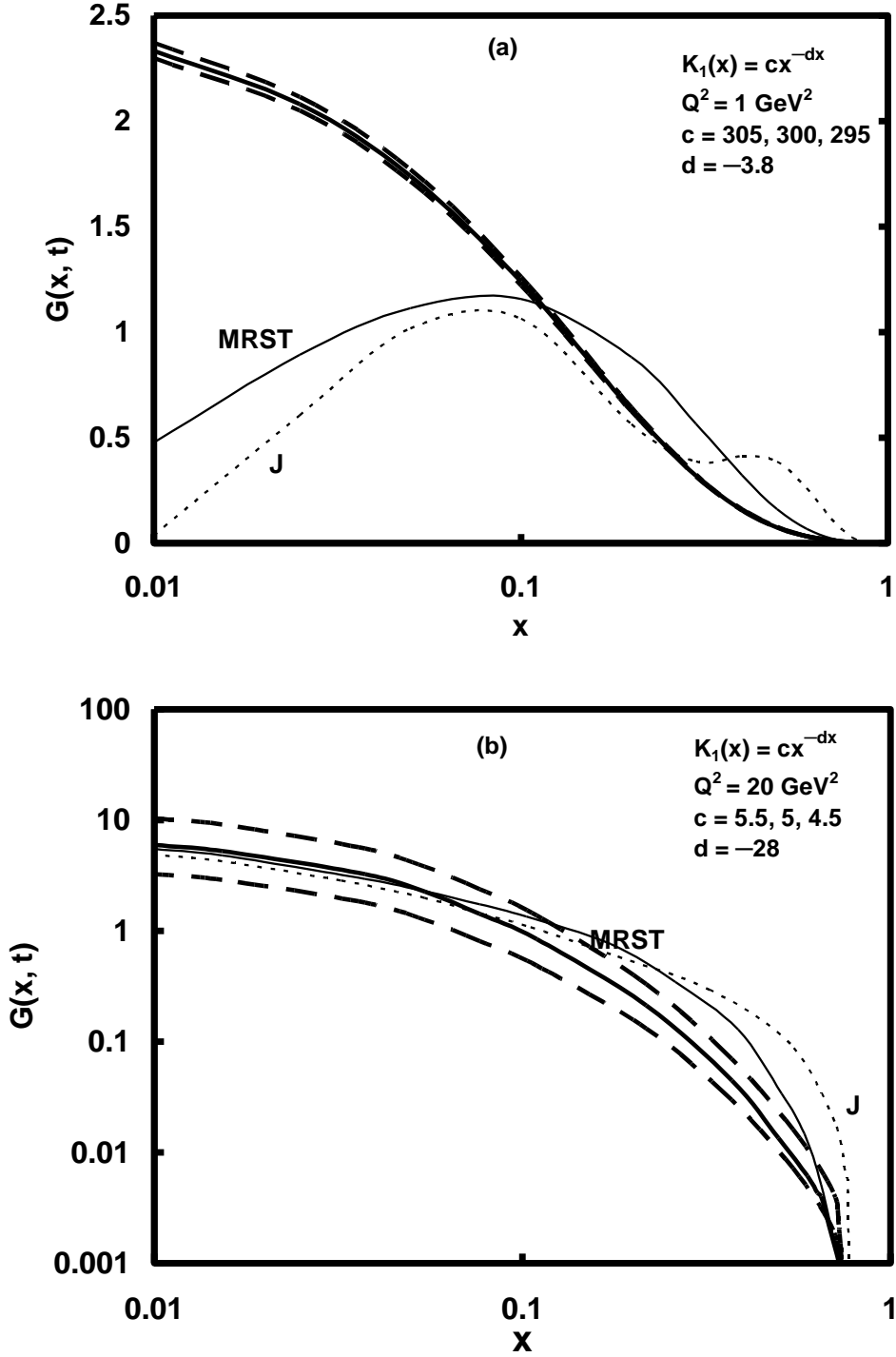

**Figure 5(a-b):** Results of *x*-distribution of gluon structure functions $G(x, t)$ for $K_1(x) = ce^{-dx}$, where '*c*' and '*d*' are constants for representative values of $Q^2$ given in each figure, and compare them with recent global parameterizations for *y* minimum in the relation $\beta = \alpha^y$. In the same figures we present the sensitivity of our results for different values of '*c*' by thick dashed lines at fixed value '*d*'. Here we take $d = -3.8$ in figure 5(a) and $d = -28$ in figure 5(b).

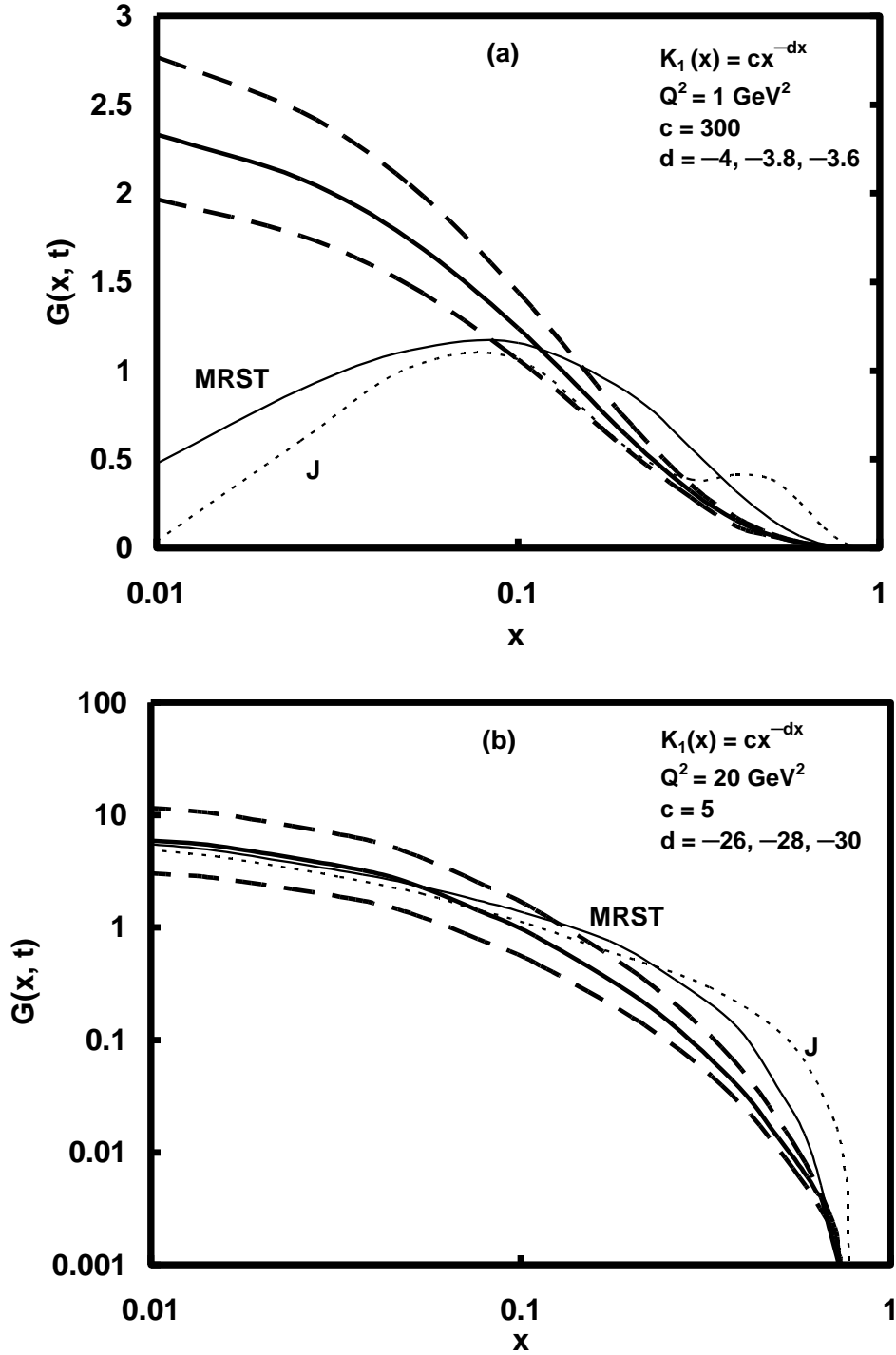

**Figure 6(a-b):** Sensitivity of our results for different values of '*d*' at fixed value of '*c*'. Here we take $c = 300$ in figure 6(a) and $c = 5$ in figure 6(b).

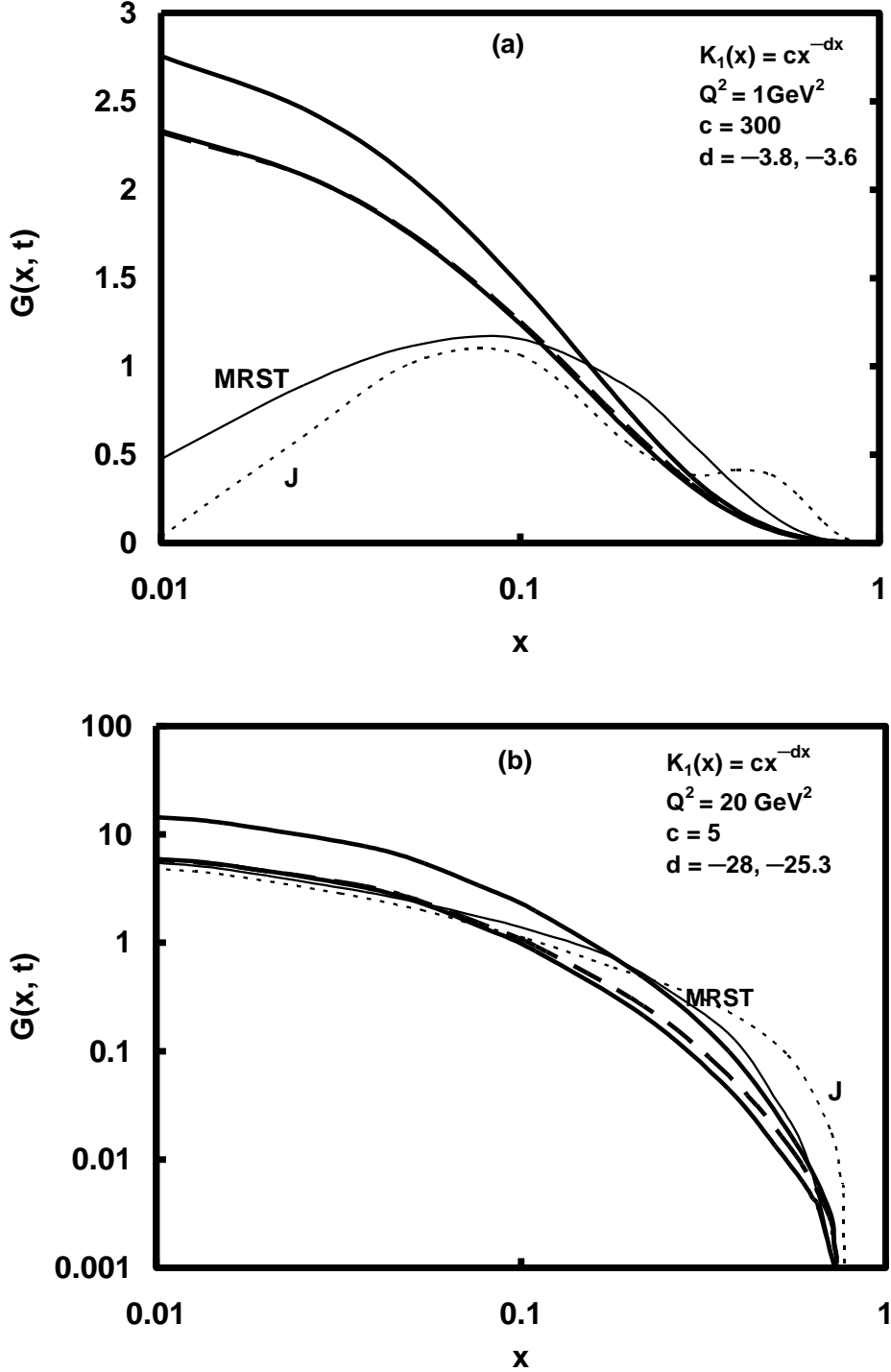

**Figure 7(a-b):** Results of $x$-evolution of gluon structure function $G(x, t)$ for $K_1(x) = ce^{-dx}$ for $y$ minimum (lower thick solid lines) and maximum (upper thick solid lines) in the relation $\beta = \alpha^y$ at same parameter values $c = 300$, $d = -3.8$ in figure 7(a) and $c = 5$, $d = -28$ in figure 7(b) and for representative values of $Q^2$ given in each figure, and compare them with recent global parameterizations. Result of $x$-evolution of gluon structure function, for $y$ maximum (long dashed lines) coincide with result of $x$-evolution of gluon structure function for $y$ minimum (lower thick solid lines) when $c = 300$, $d = -3.6$ in figure 7(a) and $c = 5$, $d = -25.3$ in figure 7(b).

In figure 6(a-b), we present the sensitivity of our results for different values of '$d$' at fixed value of '$c$'. Here we take $c = 300$ in figure 6(a) and $c = 5$ in figure 6(b). We observe that agreement of the results (thick solid lines) with parameterizations is good in figure 6(a) at $d = -3.8$, and 6(b) at $d = -28$. If value of '$d$' is increased or decreased, the curve goes downward or upward direction in figure 6(a) and if value of '$d$' is increased or decreased the curve goes upward or downward direction in figure 6(b) . But the nature of the curves is similar in both cases.

In figure 7(a-b), we present our results of $x$-evolution of gluon structure function $G(x, t)$ for $K_1(x) = ce^{-dx}$ for $y$ minimum (lower thick solid lines) and maximum (upper thick solid lines) in the relation $\beta = \alpha^y$ at same parameter values $c = 300$, $d = -3.8$ in figure 7(a) and $c = 5$, $d = -28$ in figure 7(b) and for representative values of $Q^2$ given in each figure, and compare them with recent global parameterizations [7]. We observe that result of $x$-evolution of gluon structure function, for $y$ maximum (long dashed lines) coincide with result of $x$-evolution of gluon structure function for $y$ minimum (lower thick solid lines) when $c = 300$, $d = -3.6$ in figure 7(a) and $c = 5$, $d = -25.3$ in figure 7(b). That means if $y$ varies from minimum to maximum, then value of parameter '$d$' varies from -3.8 to -3.6 in figure 7(a) and from -28 to -25.3 in figure 7(b). In these cases, value of parameter '$c$' remains constant. It is to be noted that agreement of the results with parameterization is found to be very poor for any constant value of $K_1(x)$. Therefore, we do not present our result of $x$-distribution at $K_1(x) = $ constant. Moreover, in general, the agreement of our results with the parameterization at small-$x$ is poor for low-$Q^2$ value and excellent for high-$Q^2$ value which is quite expected.

From our above discussion, it has been observed that though we can derive a unique $t$-evolution for gluon structure function in LO, yet we can not establish a unique $x$-evolution for gluon structure function in LO. $K_1(x)$, the relation between singlet and gluon structure functions, may be in the forms of a constant, an exponential function of $x$ or a power in $x$ and they can equally produce required $x$-distribution of gluon structure functions. But unlike many parameter arbitrary input $x$-distribution functions generally used in the literature, our method requires only one or two such parameters. On the other hand, The explicit form of $K_1(x)$ can actually be obtained only by solving coupled DGLAP evolution equations for singlet and gluon structure functions, and works are going on in this regard.

**References**

[1] R Rajkhowa and J K Sarma *Indian J. Phys.* **78A** 367 (2004).

[2] R Rajkhowa and J K Sarma, Particular solution of DGLAP evolution equation in next-to-leading order and structure functions at low-x, *Indian J. Phys* (in press).

[3] G Altarelli and G Parisi *Nucl. Phys* **B 126** 298 (1977).

[4] G Altarelli *Phy. Rep*. **81** 1 (1981).

[5] V N Gribov and L N Lipatov *Sov. J. Nucl. Phys*. **20** 94 (1975).

[6] Y L Dokshitzer *Sov.Phy. JETP* **46** 641 (1977).

[7] A D Martin et. al. hep-ph / 0110215 (2001).

[8] A D Martin et. al. *RAL-94-055* (1994).

[9] L F Abbott, W B Atwood and R M Barnett *Phys. Rev*. **D22** 582 (1980).

[10]. I S Gradshteyn and I M Ryzhik, *Tables of Integrals, Series and Products* Alen Jeffrey (ed) (New York: Academic) (1965).

[11] F Ayres (Jr.) *Differential Equations* (Schaum's Outline Series) (New York: McGraw-Hill) (1952).

[12] D K Choudhury and J K Sarma *Pramana-J. Phys*. **39** 273 (1992).

[13] J K Sarma, D K Choudhury and G K Medhi *Phys. Lett*. **B403** 139 (1997).

[14] J K Sarma and B Das *Phy. Lett* **B304** 323 (1993).

[15] F H Miller *Partial Differential Equation* (New York: John and Willey) (1960).

[16] F Halzen and A D Martin *Quarks and Leptons: An Introductory Course in Modern Particle Physics* (New York: John and Wiley) (1984).

[17] P D B Collins, A D Martin and R H Dalitz *Hadron Interactions* (Bristol: Adam Hilger Ltd.) (1984).

[18] B Badelek et. al. *Small-x physics* **DESY 91-124** October (1991).